\documentclass[aps,preprint,showpacs]{revtex4}

\usepackage{epsfig}

\newcommand{\beq}{\begin{equation}}
\newcommand{\eeq}{\end{equation}}
\newcommand{\beqarray}{\begin{eqnarray}}
\newcommand{\eeqarray}{\end{eqnarray}}
\newcommand{\sgn}[2][]{\ensuremath{\text{sgn}_{#1}(#2)}} 
\newcommand{\Kf}[1][\nu\sigma]{\ensuremath{\hat{F}_{#1}}} 
\newcommand{\Kfdag}[1][\nu\sigma]{\ensuremath{\hat{F}^{\dagger}_{#1}}} 
\newcommand{\com}[2]{\ensuremath{\left[#1,#2\right]_{-}}} 
\newcommand{\tfrac}{\textstyle\frac}
\newcommand{\half}{\ensuremath{\tfrac{1}{2}}}
\newcommand{\vF}{\ensuremath{v_{F}}} 
\newcommand{\kF}{\ensuremath{k_{F}}} 
\newcommand{\fdag}[1][j\sigma]{\ensuremath{f^{\dagger}_{#1}}} 
\newcommand{\Hc}{\ensuremath{\mbox{H.c.}}} 
\newcommand{\Ham}[1][]{\ensuremath{{\cal{H}}_{\text{\tiny{#1}}}}} 
\newcommand{\KD}[2]{\ensuremath{\delta_{#1,#2}}} 
\newcommand{\CO}[2][\alpha]{\ensuremath{\Lambda^{#2}_{#1}(k)}} 
\newcommand{\notag}{\nonumber}
\newcommand{\eq}[1]{Eq.~(\ref{#1})} 
\newcommand{\fig}[1]{Fig.~(\ref{#1})} 

\begin{document}

\title{Bosonization solution of the Falicov-Kimball model} 
\author{P. M. R. Brydon and M. Gul\'{a}csi} 
\affiliation{
Department of Theoretical Physics, 
Institute of Advanced Studies, 
The Australian National University, Canberra, ACT 0200, Australia}

\date{\today}

\begin{abstract}
We use a novel approach to analyze the one-dimensional spinless
Falicov-Kimball model. We derive a simple effective model for the
occupation of the localized orbitals which clearly reveals the origin
of the known ordering. Our study is extended to a quantum model with
hybridization between the localized and itinerant states: we find a
crossover between the well-known weak- and strong-coupling
behaviour. The existence of electronic polarons at intermediate
coupling is confirmed. A phase diagram is presented and discussed in
detail.  
\end{abstract}

\pacs{71.10.Fd, 71.28.+d, 71.27.+a, 71.30.+h}
\maketitle

The Falicov-Kimball model (FKM) is one of the simplest yet most
versatile models of strongly correlated electron systems. It describes
a band of conduction ($c$-) electrons interacting via a repulsive
contact potential $G$ with localized $f$-electrons of energy
$\epsilon_{f}$. Originally conceived of as a model for valence
transitions (VTs)~\cite{FKMoriginal}, the FKM is today usually
regarded as a simple model of a binary alloy, the so-called
crystallization problem (CP)~\cite{KL86}. The CP assumes fixed $c$-
and $f$-electron populations, examining the diagonal LRO (or
``crystalline order'') adopted by the $f$-electrons in the ground
state. In spite of detailed numerical maps of the one-dimensional (1D)
FKM's phase diagram~\cite{crystalnumerics}, its behaviour is only
rigorously understood at weak-~\cite{crystalanalytics} and
strong-coupling~\cite{segregation}. 

A $c$-$f$ hybridization term is usually added to the FKM in order to
model mixed-valence systems~\cite{LRP81}. The presence of quantum
valence fluctuations makes this a fundamentally different problem to
the FKM: it is referred to as the extended or quantum Falicov-Kimball
Model (QFKM). In contrast to the FKM, the QFKM's behaviour remains
largely unknown. Although there is no consensus on the nature of the
ground state~\cite{epolarons,POS96}, excitonic effects are expected to 
dominate the physics and so this remains an important open question in
the theory of optical properties of strongly correlated electron systems. 

Using the well-known non-perturbative technique of bosonization, we
obtain for the first time a complete theoretical description of the 1D
FKM below half-filling. This analysis reveals the role of the
$c$-electrons in generating the observed $f$-electron order; all
aspects of the CP results are naturally explained within our
approach. We derive an effective Hamiltonian for the $f$-orbital
occupancy which accurately predicts the FKM's phase diagram. For a
small hybridization potential our approach is also suitable for the
QFKM. We present a phase diagram that interpolates between the
well-known weak- and strong-coupling limits; our work rigorously
establishes the relevance of electronic polaron effects.  

The FKM for spinless fermions has the Hamiltonian 
\beq
\Ham[FKM] = -t\sum_{\langle{i,j}\rangle}c^{\dagger}_{i}c_{j} +
\epsilon_{f}\sum_{j}n^{f}_{j} + G\sum_{j}n^{c}_{j}n^{f}_{j} \label{eq:FKM}
\eeq
We consider only repulsive potentials, $G>0$. The concentration of
electrons is fixed at  $n =
\frac{1}{N}\sum_{j}\{n^{f}_{j}+n^{c}_{j}\}<1$  
where $N$ is the number of sites. In our study of the QFKM, we adopt
the standard on-site hybridization potential $\Ham[hyb] =
V\sum_{j}\{f^{\dagger}_{j}c_{j}+\Hc\}$.

Proceeding with our bosonization solution, we linearize the
$c$-electron spectrum about the two Fermi points and define left- and
right-moving Fermion fields $c_{\nu{j}}$, $\nu=L(-)$, $R(+)$
respectively as subscript (otherwise). The chiral density operators 
$\rho_{\nu}(k)=\sum_{k'}c^{\dagger}_{\nu{k'+k}}c_{\nu{k'}}$ 
obey the standard Luttinger commutators
$\com{\rho_{\nu}(k)}{\rho_{\nu'}(k')} =
\KD{\nu}{\nu'}\KD{k}{-k'}{\nu{kL}}/{2\pi}$ for a system of size
$L\gg{a}$, the lattice constant. The dual Bose fields are constructed
in terms of the $\rho_{\nu}(k)$~\cite{H81}: 
\beqarray
\phi(x_{j}) &=&
-i\sum_{\nu}\sum_{k\neq0}\frac{\pi}{kL}\rho_{\nu}(k)\CO{}{e}^{ikx_j}
\label{eq:Bfieldphi} \\
\theta(x_{j}) &=&
i\sum_{\nu}\sum_{k\neq0}\nu\frac{\pi}{kL}\rho_{\nu}(k)\CO{}{e}^{ikx_j} \label{eq:Bfieldtheta}
\eeqarray
The cut-off function $\CO{}$ appearing in \eq{eq:Bfieldphi} and
\eq{eq:Bfieldtheta} satisfies the conditions $\CO{}\approx1$ for
$|k|<\frac{\pi}{\alpha}$ and $\CO{}\approx0$
otherwise. $\CO{}$~enforces the finite minimum wavelength $\alpha>a$
of the bosonic density fluctuations $\rho_{\nu}(k)$: taking into
account this wavelength limit is essential for preserving the lattice
structure of the FKM~\cite{G04}. Since the Bose fields cannot
`resolve' distances less than $\alpha$, we find that the usual field
commutators are `smeared' over this length scale:
\beqarray
\com{\phi(x_{j})}{\theta(x_{j'})} &=&
\frac{i\pi}{2}{\sgn[\alpha]{x_{j'}-x_{j}}} \\
\com{\partial_{x}\phi(x_{j})}{\theta(x_{j'})} &=&
-i\pi\delta_{\alpha}(x_{j'}-x_{j}) 
\eeqarray
Here $\sgn[\alpha]{x}$ and $\delta_{\alpha}(x)$ are the $\alpha$-smeared
sign and Dirac delta functions respectively~\cite{smeared}.

The chiral Fermions are represented in terms of the Bose fields by the
Mandelstram identity 
$c_{\nu{j}} =
\sqrt{{A}a/\alpha}\Kf[\nu]\exp\left(-i\nu\left[\phi(x_j)-\nu\theta(x_j)\right]\right)$.
The dimensionless normalization constant $A$ is dependent upon the
form of $\CO{}$; the Klein factors $\Kf[\nu]$ are ``ladder operators'' 
between subspaces of differing $c$-electron number. The bosonic form
of the number operators is given by 
\beq
n^{c}_{j} = n^{c}_{0}-\frac{a}{\pi}\partial_{x}\phi(x_j) +
\frac{Aa}{\alpha}\sum_{\nu}\Kfdag[\nu]\Kf[-\nu]e^{i2\nu\phi(x_j)}e^{-i2\nu{k}_{F}x_{j}} \label{eq:B:nb_identity}
\eeq
The first term on the RHS, $n^{c}_{0}$, is the non-interacting
$c$-electron concentration; the second term is the forward-scattering
density fluctuation; the third term is the first order backscattering
correction. Higher order corrections are usually neglected.

Our bosonization analysis requires that $n^{c}_{0}\neq0$. For the sake
of brevity, here we assume the symmetric case with equal $c$- and
$f$-populations in the non-interacting limit,
i.e. $n^{c}_{0}=n^{f}_{0}=\half{n}$. For the VT problem, this pins the  
Fermi level at $\epsilon_{f}=-2t\cos(\kF{a})$, $\kF=\pi{n}/2a$ is the
Fermi momentum (note that $\epsilon_{f}$ is irrelevant to the
CP). Using the electron concentration condition we also re-write the
Coulomb interaction up to a constant: $G\sum_{j}n^{c}_{j}n^{f}_{j} = 
G\sum_{j}n^{c}_{j}\left(n^{f}_{j}-\half\right)
-\half{G}\sum_{j}n^{f}_{j}$. 
It is now a simple matter to obtain the bosonized form of the
Hamiltonian by substituting these identities into \eq{eq:FKM}.

We apply a shift transformation on the $c$-electron bosonic fields
$\hat{U} =
\exp\left\{i\frac{Ga}{\pi\vF}\sum_{j}(n^{f}_{j}-\half)\theta(x_j)\right\}$
where $\vF=2ta\sin(\kF{a})$ is the $c$-electron Fermi velocity.
This transformation rotates the basis of the Hilbert space so that the 
$c$-electrons are explicitly coupled to the $f$-orbitals. Introducing
a pseudospin-$\half$ representation for the $f$-electron occupation at
site $j$, $n^{f}_{j}-\half = \tau^{z}_{j}$, we write the transformed
Hamiltonian 
\beqarray
{\hat{U}}^{\dagger}\Ham[FKM]{\hat{U}}&=&\frac{\vF{a}}{2\pi}\sum_{j}\left\{\left(\partial_{x}\phi(x_j)\right)^{2}+\left(\partial_{x}\theta(x_j)\right)^{2}\right\}+{\half}{G}\left(n-1\right)\sum_{j}n^{f}_{j}-{\frac{G^2a^2}{2\pi\vF}\sum_{j,j'}\tau^{z}_{j}\delta_{\alpha}(x_j-x_{j'})\tau^{z}_{j'}}
\notag \\
&&-\frac{2GAa}{\alpha}\sum_{j}\tau^{z}_{j}\cos\left(2\left\{\phi(x_j)-
{\cal{K}}_{\alpha}(x_j)
-\left[\kF+\frac{\pi}{2a}\right]x_j\right\}\right)
\label{eq:EPH:CTHam} 
\eeqarray
where ${\cal{K}}_{\alpha}(x_j) =
{\cal{S}}_{\alpha}(x_j)+{\cal{L}}_{\alpha}(x_j)$,
${\cal{S}}_{\alpha}(x_j) =
\frac{\pi}{2}\sum_{n=1}^{\infty}[\sgn[\alpha]{x_{j+n}-x_j}-1]\left(\tau^{z}_{j+n}-\tau^{z}_{j-n+1}\right)$
and  
${\cal{L}}_{\alpha}(x_j) =
\frac{\pi}{2}\left(\frac{Ga}{\pi\vF}-1\right)\sum_{n=1}^{\infty}\sgn[\alpha]{x_{j+n}-x_j}\left(\tau^{z}_{j+n}-\tau^{z}_{j-n}\right)$.
${\cal{S}}_{\alpha}(x_j)$ is a measure of the short-range behaviour of the
pseudospins, subtracting the $\tau^{z}$ within $\alpha$ to the
left of $x_{j}$ from the $\tau^{z}$ within $\alpha$ to the right;
${\cal{L}}_{\alpha}(x_j)$ probes the long-range behaviour, subtracting
the $\tau^{z}$ more than $\alpha$ to the left of $x_j$ from the
$\tau^{z}$ more than $\alpha$ to the right~\cite{stringop}. Note that
we retain all terms produced by the canonical transform
in~\eq{eq:EPH:CTHam}. 

The removal of the forward-scattering Coulomb interaction introduces
the important new term 
\beq
\Ham[int] = -\frac{G^2a^2}{2\pi\vF}\sum_{j,j'}\tau^{z}_{j}\delta_{\alpha}(x_j-x_{j'})\tau^{z}_{j'} \label{eq:EPH:DE}
\eeq
The physical origin of this interaction is the delocalization of the
$c$-electrons below scales less than $\alpha>a$. A $c$-electron spread
over more than a single lattice site carries the same charge over
these sites: due to the Coulomb interaction $G$, this will favour
empty underlying $f$-orbitals. Although such ``segregation'' is
known to occur in the CP~\cite{crystalnumerics,segregation}, the
derivation of the responsible effective interaction is a new
achievement in the history of the model. The $\alpha$-smeared
$\delta$-function in \eq{eq:EPH:DE} makes it clear that the
interaction is short-ranged; we may therefore approximate the
interaction as nearest-neighbour only,
$\Ham[int]\approx-{\cal{J}}\sum_{j}\tau^{z}_{j}\tau^{z}_{j+1}$, 
${\cal{J}} = G^2a^2\delta_{\alpha}(a)/\pi\vF$.

The other terms in~\eq{eq:EPH:CTHam} involving the pseudospins are a
constant and a site-dependent longitudinal field. The former is
important only in the VT: together with the first term 
in~\eq{eq:EPH:CTHam} this determines the distribution of the total
electron population across the two orbitals. The site-dependent field
arises from the $2\kF$-backscattering of the $c$-electrons off the
$f$-orbitals. This is responsible for the observed LRO phases, in
close analogy to the Peierls state~\cite{crystalanalytics}. It is
clear from~\eq{eq:EPH:CTHam} that the LRO dominates the segregation at
weak-coupling $G\ll{t}$; increasing $G$, however,~\eq{eq:EPH:DE}
eventually causes the system to segregate (the SEG phase).

The competition between the SEG and LRO phases can be simply studied
within the framework of our bosonization approach: since the only
coupling in~\eq{eq:EPH:CTHam} between the Bose fields and the
$\tau$-pseudospins is in the backscattering term, by replacing $\phi$
with a suitably chosen expectation value we obtain an effective
Hamiltonian $\Ham[eff]$ for the $f$-occupation {\it{only}}. Exact
diagonalization calculations on $3200$-site chains reveals that  
within the SEG phase the $c$-electrons are at their Luttinger liquid
fixed point for all $G$~\cite{F03}. The choice
$\langle\phi(x_j)\rangle=0$ is therefore valid across the phase
diagram~\cite{G04}. Substituting this into~\eq{eq:EPH:CTHam}, we find
\beqarray
\Ham[eff] &=&
-{\cal{J}}\sum_{j}\tau^{z}_{j}\tau^{z}_{j+1} \notag \\
&& -\frac{2GAa}{\alpha}\sum_{j}\tau^{z}_{j}\cos\left(2\left\{
{\cal{K}}_{\alpha}(x_j)
+\left[{\kF}+{\tfrac{\pi}{2a}}\right]x_j\right\}\right) \notag\\
\label{eq:EPH:effHam} 
\eeqarray
Below we present a detailed analysis of $\Ham[eff]$.

Since the $c$- and $f$-populations are fixed in the CP,
\eq{eq:EPH:effHam} by itself fully describes the system close to the
SEG-LRO boundary. To determine this line the site-dependence of the
longitudinal field must be known; in particular, we examine  
${\cal{S}}_{\alpha}(x_j)$ and ${\cal{L}}_{\alpha}(x_j)$. 
Within the SEG phase, the $f$-electrons are arranged into a single
block~\cite{segregation}. The magnitude of ${\cal{L}}_{\alpha}(x_j)$
reaches a maximum at the edge of this block, increasingly linearly as
the edge is approached from either side. In contrast,
${\cal{S}}_{\alpha}(x_j)$ vanishes everywhere except in the vicinity
of this edge. The longitudinal field thus has the general form
$h^{z}_{j}\propto\cos(\omega{j}+\phi)$, $\omega$ and $\phi$ constants.  

\begin{figure}[t]
\includegraphics[width=6.5cm]{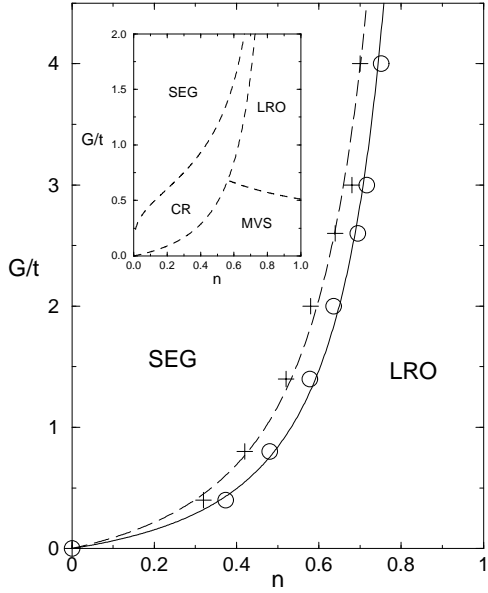} 
\caption{\label{fig:PD}Ground-state phase diagram for the symmetric
FKM below half-filling for the CP (dashed line) and VT problem (solid
line). The numerical data points (circles for VT, plus sign for CP)
are due to Gajek {\it{et al.}}~\cite{crystalnumerics}.  
Inset: the phase diagram for the QFKM. Phases are as described in text.}
\end{figure}

The periodicity ($\omega$) of $h^{z}_{j}$ determines the
critical Ising coupling~\cite{S93}: for $G\ll{t}$ we find the critical 
line $G_{c}a={\vF}\sqrt{2\pi{A}/\alpha\delta_{\alpha}(a)}$.  
The $f$-electron configuration acts as an applied scattering potential
$G\langle\tau^{z}_{j}\rangle$ for the $c$-electrons~\cite{F03}. Below
the average interparticle separation $\sim\kF^{-1}$ the $c$-electrons
scatter independently: for $n\ll{1}$ the low-energy ($E=0$) wavefunctions
$\psi$ therefore obey the Schr\"{o}dinger equation
$\partial^{2}_{x}\psi(x)=Gm\langle{\tau^{z}_{x}}\rangle\psi(x)$ in the
continuum limit ($m$ is the bare electron mass). From elementary
quantum mechanics we find plane wave solutions over unoccupied
orbitals ($\langle\tau^{z}_{x}\rangle=-\half$) and exponentially
decaying solutions with characteristic length $\zeta\propto\sqrt{t/G}$
over occupied orbitals
($\langle\tau^{z}_{x}\rangle=\half$)~\cite{Schiff}. $\zeta$ 
clearly corresponds to $\alpha$ in the boson theory; for small $G_{c}$
Taylor-expanding $\delta_{\alpha}(a)$ in the critical line equation
gives $G_{c}a/\vF=\sqrt{2\pi^2{A}}(1+a^2/2\alpha^2+\ldots)$,
implying a linear relationship between $G_{c}a/\vF$ and
$G_{c}$. This is confirmed by examining the numerical data of
Gajek \emph{et al.}, with our fit giving $A=1/32\pi^2$,
$\alpha=a/2\sqrt{G_{c}}$, and a critical line $G_{c}/t = 
0.5\sin(n\pi/2)/(1-\sin(n\pi/2))$ as plotted in 
\fig{fig:PD}. This is the first analytic expression for the SEG-LRO
boundary valid at weak coupling.

Moving on to the VT problem, it is clear from \eq{eq:EPH:CTHam} that
the Coulomb interaction will shift the $f$-level downwards, emptying
the $c$-band. The depletion of the $c$-electron population causes
segregation to occur at a smaller value of $G$ than in the CP; the new
boundary can be calculated self-consistently from the available
numerical results~\cite{crystalnumerics}. Using the same curve-fitting
technique as for the CP, we find excellent agreement with the data for
$G_c/t=0.35\sin(n\pi/2)/[1-\sin(n\pi/2)]$
($A\approx1.55\times10^{-3}$, $\alpha\approx0.418a/\sqrt{G_{c}}$), see
\fig{fig:PD}. Further increasing $G$ above the critical value $G/t 
= 4(1-\cos(n\pi/2))/(1-n)$ the $f$-level lies below the $c$-electron
band edge and so all electrons have $f$-character. The absence of
$c$-electrons to mediate the segregating interaction implies that here
any $f$-configuration is the ground state.   

We extend our analysis to the physically interesting limit $V{\ll}t$
of the QFKM. Bosonizing and carrying out the canonical transform, the
hybridization adds the term  
\beq
{\hat{U}}^{\dagger}\Ham[hyb]\hat{U} =
4V\sqrt{\frac{Aa}{\alpha}}\sum_{j}{\tau}^{x}_{j}\cos\left(
{\cal{K}}_{\alpha}(x_j)
+ \left[\kF + {\tfrac{\pi}{2a}}\right]x_j\right) \label{eq:hyb}
\eeq
to the effective pseudospin Hamiltonian \eq{eq:EPH:effHam}. We have used a
generalized Jordan-Wigner transformation to combine the $f$-electron
operators and the Klein factors into pseudospins
$\tau^{+}_{j}=\fdag[j]\Kf[\nu]e^{-i\nu\pi{x_{j}}/2a}{\exp\left({-i\frac{\nu\pi}{2}\sum_{j'}\sgn{x_{j'}-x_j}\left(n^{f}_{j'}-\half\right)}\right)}$. 
The resulting effective pseudospin Hamiltonian is easily recognized as
the Ising model in site-dependent longitudinal ($h^{z}_{j}$) and
transverse ($h^{x}_{j}$) fields.

Although we are not aware of any systematic study of this Hamiltonian,
we can nevertheless deduce important aspects of its physical behaviour.
The two extreme limits $G{\gg}t$ and $G{\ll}t$ are easiest to
discuss. In the former case the presence of $V\ll{G}$ does not
radically modify the VT physics. Where the longitudinal field
dominates, we can discard the Ising interaction so that the
pseudospins track the longitudinal and transverse fields; clearly the
transverse field will only be important where the longitudinal field
is small. This will ``smear'' the crystalline arrangement, but leaves
the basic picture of a density wave intact. Where the Ising
interaction dominates, the hybridization will stabilize
the SEG phase as quantum fluctuations maintain a non-zero
$c$-electron population.  

For $G\ll{t}$ the quantum fluctuations produced by
the hybridization dominate, fundamentally altering the
behaviour of the system. Discarding the Ising interaction which is
irrelevant for most of this regime, we arrive at the same trivial
model of pseudospins tracking longitudinal and transverse fields as in
the $G\gg{t}$ LRO phase. Here, however, the transverse field
is generally much larger than the longitudinal field: the
pseudospins will be mostly ordered along the $x$-axis with
$\langle\tau^{x}_{j}\rangle\approx\half\sgn{h^{x}_{j}}$. In terms of the
$f$-occupation, this implies that
$\langle{f^{\dagger}_{j}c_{j}}\rangle\neq0$, with intermediate 
occupation of each $f$-orbital
$0<\langle{n^{f}_{j}}\rangle\approx{\half{n}}$. The longitudinal field  
produces only small deviations from this homogeneous distribution of
electron density. This is clearly a mixed-valence state (MVS). 

Our pseudospin model allows us to connect these two extreme limits for
the first time: for $V=0.05t$ we present the QFKM's phase diagram as
the inset in \fig{fig:PD}. Close to half-filling the Ising interaction
is never dominant so the model of $\tau$-pseudospins tracking the
longitudinal and transverse fields is valid for any
$G$. In~\fig{fig:PD} we define an MVS-LRO boundary by equating the
magnitude of the transverse and longitudinal fields. For small $n$,
however, the longitudinal field is negligible; discarding this term
from $\Ham[eff]$ we find a transverse field Ising model (TFIM)
describes the $f$-orbital occupation. Although the TFIM has an
order-disorder transition~\cite{P79}, the shifting of the $f$-level
lifts the system from criticality and so there is a crossover regime
(CR) between the SEG and MVS phases.    

The details of the CR physics are controlled by the site-dependence
of the transverse field \eq{eq:hyb}. Near the SEG phase we can
use the same values for ${\cal{S}}_{\alpha}(x_j)$ and
${\cal{L}}_{\alpha}(x_j)$ as in the CP analysis; for MVS behaviour
both these objects vanish. We thus assume a periodic
transverse field within the CR. In general, the periodicity is
incommensurate with the lattice: numerical studies~\cite{Satija} reveal
this to be qualitatively identical to a random variation of the
transverse field. The random TFIM has been studied in depth by
Fisher~\cite{F95} using a real-space renormalization group
treatment. To relate Fisher's results to the CR, we assume an 
effective Hamiltonian $\Ham[CR] =
-{\cal{J}}\sum_{j}\tau^{z}_{j}\tau^{z}_{j+1}+\sum_{j}h^{x}_{j}\tau^{x}_{j}$  
where the values of $h^{x}_{j}$ are drawn randomly from the cosine
distribution $\rho(h)dh = 
(C\pi)^{-1}\sqrt{1-(h/C)^2}$, $C=4V\sqrt{Aa/\alpha}$. 
For the CR, $A$ and $\alpha$ are as in the SEG phase; we therefore
use the values from the classical ($V=0$) VT problem.

Lowering $G$ from the SEG phase, the CR is reached at
${\cal{J}}_{c1}=\max{|h^{x}_{j}|}$: for ${\cal{J}}<{\cal{J}}_{c1}$ the
Ising coupling is not greater than the magnitude of the transverse
field \emph{everywhere} on the lattice. The rare regions where 
$|h^{x}_{j}|>{\cal{J}}$ breaks the SEG ground state up into randomly
distributed large clusters of zero and full $f$-occupancy. The system
nevertheless retains the character of the SEG phase so long as
${\cal{J}}>{\cal{J}}_{c2}=\overline{|h^{x}_{j}|}$, the average value
of $|h^{x}_{j}|$ across the lattice. This is reflected by the mean
$f-f$ correlations $\overline{\langle{n^{f}_{j}n^{f}_{j+x}}\rangle}$
which decay exponentially to $(n^{f})^2>(n^{f}_{0})^2$ with
correlation length $\xi\sim{\exp\{-2[\ln({\cal{J}}_{c2}/{\cal{J}})]^2\}}$.

Decreasing ${\cal{J}}$ below ${\cal{J}}_{c2}$, the character of the
CR reverses, as here ${\cal{J}}<|h^{x}_{j}|$ over most of the
lattice. Of the SEG state, only rare clusters remain, embedded in a
valence-fluctuating background. This MVS-like region of the
CR is characterized by a different functional form for the correlation
length $\xi\sim[\ln({\cal{J}}_{c2}/{\cal{J}})]^{-2}$; the correlation
functions decay exponentially to $(n^{f}_{0})^2$. The CR persists
until ${\cal{J}}$ becomes comparable to the longitudinal field in
\eq{eq:EPH:effHam}: the system is then best described by the small-$G$
MVS. In \fig{fig:PD} the upper boundary of the CR is derived from the
condition ${\cal{J}}_{c1}=\max{|h^{x}_{j}|}$ and is given by $G_{c1}=
2[\{V\vF\pi/a\delta_{\alpha}(a)\}\sqrt{A/a\alpha}]^{1/2}$;
the lower boundary is the critical line $G_{c}$ as in the
classical VT problem. We have extended the curve $G_{c1}(n)$ to $G>t$
using the VT result for $\alpha$. 

The distinguishing feature of the CR are the randomly distributed
clusters of integral $f$-valence. These clusters occur where the local
$c$-electron density departs from the MVS average, determining the
occupation of the underlying $f$-orbitals via the forward-scattering
Coulomb interaction [i.e.~\eq{eq:EPH:DE}]. This coupling of
the $c$- and $f$-electron densities is recognizable as a Toyozawa
electronic polaron~\cite{T54}. Electronic polarons in the QFKM were
studied by Liu, who proposed that they appear intermediate between
mixed- and integral-valence phases~\cite{epolarons}. Since
bosonization treats the forward-scattering exactly~\cite{H81}, our
calculation rigorously confirms Liu's scenario.

In summary, we have presented the results of a novel theoretical
study of the 1D FKM. Using a non-perturbative approach, we have
uncovered the physical mechanisms responsible for the known
$f$-electron ordering. We derive an effective Hamiltonian for the
occupancy of the $f$-orbitals which predicts the SEG-LRO
transition. We also study the QFKM for small hybridization
$V\ll{t}$. For the first time we can accurately interpolate between
the well-known CP and MVS limits: at intermediate coupling we find a
crossover regime where electronic polaron effects are of
importance. An extensive discussion of this system is to follow in a
later paper~\cite{BG05}.    

PMRB acknowledges useful discussions with C. D. Batista and
J. E. Gubernatis.


\begin{thebibliography}{99}
\bibitem{FKMoriginal}L. M. Falicov and J. C. Kimball,
Phys. Rev. Lett. {\bf{22}}, 997 (1969).

\bibitem{KL86}T. Kennedy and E. H. Lieb, Physica A {\bf{138}}, 320 (1986).

\bibitem{crystalnumerics}Z. Gajek, J. J\c{e}drzejewski and R. Lema\'{n}ski,
Physica A {\bf{223}}, 175 (1996); Phase Transit. {\bf{57}}, 139 (1996).

\bibitem{crystalanalytics}J. K. Freericks, Ch. Gruber and
N. Macris, Phys. Rev. B {\bf{53}}, 16189 (1996).

\bibitem{segregation}P. Lemberger, J. Phys. A:
Math. Gen. {\bf{25}}, 715 (1992).

\bibitem{LRP81}J. M. Lawrence, P. S. Riseborough and R. D. Parks,
Rep. Prog. Phys. {\bf{44}}, 1 (1981).

\bibitem{epolarons}S. H. Liu, Phys. Rev. Lett. {\bf{58}}, 2706 (1987).

\bibitem{POS96}T. Portengen, Th. \"{O}streich and L. J. Sham,
Phys. Rev. Lett. {\bf{76}}, 3384 (1996).

\bibitem{H81}F. D. M. Haldane, J. Phys. C {\bf{14}}, 2585 (1981).

\bibitem{G04}M. Gul\'{a}csi, Adv. Phys. {\bf{53}}, 769 (2004).

\bibitem{smeared}We use $\CO{}=e^{-\alpha|k|}$ and so we have 
$\delta_{\alpha}(x) = \frac{1}{\pi}\frac{\alpha}{\alpha^2+x^2}$,
$\sgn[\alpha]{x} = \frac{2}{\pi}\arctan(\frac{x}{\alpha})$. 

\bibitem{stringop}Using spin-$\half$ identities the
    argument of the cosine is considerably simplified. We retain
    the general form, however, as these cancellations cannot be
    made when hybridization is present.

\bibitem{F03}P. Farka\v{s}ovsk\'{y}, Int. J. Mod. Phys. B {\bf{17}},
  4897 (2003).

\bibitem{S93}C. Sire, Int. J. Mod. Phys. B {\bf{7}}, 1551 (1993). 

\bibitem{Schiff}L. I. Schiff, {\it{Quantum Mechanics}} (McGraw-Hill,
  New York, 1955).

\bibitem{P79}P. Pfeuty, Phys. Lett. A {\bf{72}}, 245 (1979).

\bibitem{Satija}I. I. Satija, Phys. Rev. B {\bf{41}}, 7235 (1990).

\bibitem{F95}D. S. Fisher, Phys. Rev. B {\bf{51}}, 6411 (1995).

\bibitem{T54}Y. Toyozawa, Prog. Theor. Phys. {\bf{12}}, 421 (1954).

\bibitem{BG05}P. M. R. Brydon and M. Gul\'{a}csi (in preparation). 
\end{thebibliography}
\end{document}